\documentclass[twocolumn,showpacs,preprintnumbers,amsmath,amssymb,prb,aps]{revtex4-1}
\usepackage{epsfig}
\usepackage{epstopdf}
\usepackage{graphicx}
\usepackage{dcolumn}
\usepackage{bm}
\usepackage{textcomp}
\usepackage{array}
\usepackage{subcaption}
\usepackage{booktabs}
\usepackage{natbib}
\usepackage{url}
\setcitestyle{numbers,square}

\begin{document}

\title{Studying the lifetime of charge and heat carriers due to intrinsic scattering mechanisms in FeVSb half-Heusler thermoelectric}
\author{Shivprasad S. Shastri}
\altaffiliation{Electronic mail: shastri1992@gmail.com}
\author{Sudhir K. Pandey}
\altaffiliation{Electronic mail: sudhir@iitmandi.ac.in}
\affiliation{School of Engineering, Indian Institute of Technology Mandi, Kamand - 175075, India}


\begin{abstract}
This work, presents a study of lifetime of carriers due to intrinsic scattering mechanisms $viz.$ electron-electron (EEI), electron-phonon (EPI) and phonon-phonon interactions (PPI) in a promising half-Heusler thermoelectric FeVSb. Using the full-$GW$ method, the effect of EEI and temperature on the valence and conduction band extrema and band gap are studied. The lifetime of carriers with temperature are estimated at these band extrema. At 300 K, estimated value of lifetime at VBM (CBM) is $\sim$1.91 x10$^{-14}s$ ($\sim$2.05 x10$^{-14}s$). The estimated ground state band gap considering EEI is $\sim$378 meV. Next, the effect of EPI on the lifetime of electrons and phonons with temperature are discussed. The comparison of two electron lifetimes suggests that EEI should be considered in transport calculations along with EPI. The average acoustic, optical and overall phonon lifetimes due to EPI are studied with temperature. Further, the effect of PPI is studied by computing average phonon lifetime for acoustic and optical phonon branches. The lifetime of the acoustic phonons are higher compared to optical phonons which indicates acoustic phonons contribute more to lattice thermal conductivity ($\kappa_{ph}$). The comparison of phonon lifetime due to EPI and PPI suggests that, above 500 K EPI is the dominant phonon scattering mechanism and cannot be ignored in $\kappa_{ph}$ calculations. Lastly, a prediction of the power factor and figure of merit of n-type and p-type FeVSb is made by considering the temperature dependent carrier lifetime for the electronic transport terms. This study shows the importance of considering EEI in electronic transport calculations and EPI in phonon transport calculations. Our study is expected to provide results to further explore the thermoelectric transport in this material.  

\end{abstract}

\maketitle

\section{Introduction} 
Thermoelectric materials are the materials in which application of temperature (electric field) gradient creates an electrical potential difference (temperature difference). This type of response to the external applied field has lead to use these materials for power generation or cooling applications utilizing the energy conversion property between heat and electricity. The efficiency of conversion from heat to electricity is connected to material specific transport coefficients via a dimensionless quantity called figure of merit $ZT$. This quantity is useful in selecting the materials for thermoelectric application \cite{compintro,tritt} . 

The figure of merit $ZT$ is related to transport terms as $ZT=S^{2}\sigma \kappa^{-1}T$ \cite{snydercomplex,mgdintro}. Here, $S$ is the Seebeck coefficient, $\sigma$ is electrical conductivity, $\kappa$ is total thermal conductivity and $T$ is absolute temperature. The $\kappa$ is the sum of electronic thermal conductivity $\kappa_{e}$ and lattice thermal conductivity $\kappa_{ph}$. Therefore, to screen a material these transport coefficients need to be theoretically calculated or experimentally measured as a function of temperature. The nature of transport coefficients are decided by the electronic and phonon spectra of the material. The dependence comes through the group velocity, specific heat, effective mass, carrier concentration (chemical potential) and relaxation time (or lifetime) of carriers. So, the theoretical or computational study of transport involves the calculation of these mentioned physical quantities \cite{ashcroft,chenreview,srivastavareview}. 

In the first-principles study, design or in the prediction of materials for thermoelectric applications, generally density functional theory (DFT) based methods are widely employed \cite{yang100hh,jesus,madsensearch}. From DFT based calculations the electronic dispersion and phonon spectrum are obtained. Once the dispersion is obtained, the quantities like group velocity, effective mass etc. can be further derived using their relation to dispersions. Temperature dependent carrier concentration can also be obtained using the distribution functions. The semi-classical or Boltzmann transport theory is the one of the most widely used tool in this direction along with DFT \cite{madsensearch, boltztrap,djsemi}. However, in this approach the situation is hindered when it comes to the calculation of lifetime of carriers, due to the non-interacting picture considered in DFT which is used to obtain the energy spectrum. Therefore, normally a constant relaxation time approximation (CRTA) is used to obtain the transport coefficients. Under this approximation, the $S$ becomes independent of lifetime, while $\sigma$ and $\kappa_{e}$ are calculated in the form $\sigma/\tau$ and $\kappa_{e}/\tau$, where $\tau$ is lifetime  \cite{boltztrap, boltztrap2, boltzwann}. So, to get the full transport terms,  a constant value for lifetime is used in the full temperature range \cite{singhcrt}, or some model for lifetime is used \cite{taumodel,zoutau}. In some cases, by fitting the available experimental $\sigma$, lifetime is obtained . These approaches may be simple and useful only in particular cases. However, they may not capture the effect of various scattering sources or temperature properly on the transport properties. Also, these approaches make fully first-principles calculation and more certain prediction of thermoelectric properties difficult.  

Along with other carrier energy dispersion derived quantities, the lifetime is an important quantity in controlling the temperature dependent transport behaviour. The lifetime of carriers in a material is decided by the sources of scattering (or interaction), strength of interaction and type of interaction. The impurities or defects act as a source of scattering (extrinsic) mainly in the lower temperature. In a perfect crystal the electron-electron interaction (EEI), electron-phonon interaction (EPI) and phonon-phonon interaction (PPI) mainly act as intrinsic sources of scattering mechanisms \cite{ashcroft}. These scattering mechanisms decide the temperature dependent behaviour of transport coefficients. An electron will have finite lifetime due to the interactions with other electrons and phonons. Similarly, phonon will have lifetime due to phonon-phonon and phonon-electron interactions. Quantification of these scattering mechanisms through lifetime or linewidth calculation helps to understand transport, identify dominant scattering source and fully first-principles prediction of thermoelectric properties. 

Recent advancements in the many-body theory and its implementation as computational methods are becoming useful to explore the electron-electron, electron-phonon or phonon-phonon interaction effects in solids \cite{gwreview,epireview,phono3py}. These methods have enabled to see the effect of the interactions on the electronic or phonon spectrum by calculation of self-energy. The imaginary part of self-energy gives a picture of strength of interaction between carriers. The electronic structure and strength of EEI in solids is normally studied  using $GW$ or DFT + dynamical mean field theory (DMFT) methods\cite{antikvan,antikiron,pdutta}. The electron-phonon matrix elements can be calculated under combined density functional perturbation theory (DFPT) and using Wannier functions for interpolation method \cite{epwtheory}. Using the interpolated electron-phonon matrix elements on finer mesh, the imaginary part of self-energy and lifetime of carriers due to EPI can be obtained. This method has been found to be increasingly applied to study the EPI in materials. Study of the PPI is important in order to understand the $\kappa_{ph}$ of thermoelectric materials. This interaction is studied by calculating phonon lifetime and imaginary part of self-energy due to PPI using the anharmonic force constant under many-body perturbation theory\cite{phono3py}. 

At the present stage, the fully first-principles calculation of thermoelectric properties considering all the above interactions may be treated as computationally demanding work. However, eventually this difficulty may be overcome, considering the increased computing power and modern codes\cite{gwreview}. Such efforts to study carrier lifetime and thermoelectric properties on selected material classes can not only improve  basic understanding of transport, but also can provide a database which can be used to create a possible lifetime model to directly apply for some material classes in the prediction or search of new thermoelectric materials thereby saving costly computations.

A number of studies have been made in predicting or understanding few aspects  of thermoelectric or transport properties on some materials considering the carrier scattering and lifetime. For instance, the thermoelectric properties studies of silicon are carried out considering the EPI and PPI by first-principles method \cite{si1,si2,si3,si4}. The mobility of n-type Si, Al and MoS$_{2}$ was calculated by W. Li considering the EPI by \textit{ab-initio} method which is in in good agreement with experiments \cite{lisi}. J. Zhou and M. Bernardi calculated electron mobility, and lifetime of electrons and phonons due to EPI in GaAs providing a scheme to study transport in polar materials \cite{gaas}. Electronic mobility of thermoelectric material n-type PbTe is calculated as a function of temperature and carrier concentration by J. Cao \textit{et al.} considering the EPI identifying the dominant scattering source \cite{npbte}. Similarly, thermoelectric parameters at 300 K in p-type PbTe is studied in Ref. \cite{ppbte} taking into account EPI. S. Li \textit{et al.} studied the electrical and thermal transport parameters, power factor and $ZT$ with change in carrier concentration at 300 K considering PPI, EPI, and phonon-dopant interaction for doped SnSe \cite{snse}. For some of half-Heusler thermoelectric materials a study has been carried out by G. Samsonidze \textit{et al.} \cite{kozinsky} proposing a new electron-phonon averaged approach (EPA) in calculating scattering rate treating EPI. The temperature dependent thermoelectric properties of p-type HfCoSb and n-type HfNiSn are calculated using this approach. A prediction of $ZT$ is made for number of half-Heuslers including FeVSb at 673 K setting value of $\kappa_{ph}$ to be 2 Wm$^{-1}$K$^{-1}$. 

Many of compounds of Heusler family have drawn attention as promising thermoelectric materials for high temperature applications due to many desirable material features \cite{hhreview1,hhreview2,hhreview3,hhreview4}. Therefore, it is important to study the different carrier scattering mechanism in this class of materials in order to further understand and predict thermoelectric properties. Considering this aspect, in our work we focus on the charge and heat carrier lifetimes due to intrinsic scattering mechanisms in one of the promising half-Heusler thermoelectric FeVSb. Firstly, the effect of EEI with temperature on the valence band maximum (VBM) and conduction band minimum (CBM) is studied under full-$GW$ method. The carrier lifetime at VBM and CBM with temperature are estimated. The change in band gap due to EEI with temperature is also discussed. Next, the effect of EPI is studied by calculating linewidth of electrons and phonons using the EPW method. The temperature dependent variation of electron and phonon lifetimes due to EPI is calculated. The variation of phonon lifetime with temperature is discussed for different branches. Similarly, the effect of PPI on the lifetime of phonons are studied by calculating branch wise lifetime.  By comparing the electron and phonon lifetime due to different scattering mechanisms, major sources of carrier scattering are identified in FeVSb. Then, the power factor and $ZT$ are estimated for p-type and n-type FeVSb using the temperature dependent carrier lifetime.


\section{Brief description of the calculation methods}
In our work mainly three different computational tools are used to study the lifetime of carriers due to intrinsic scattering mechanisms. The methods to used to study different interactions and calculate imaginary part of carrier self-energy are briefly described in this section.

In the full-$GW$ method, the spectral function $A(\mathbf{k},\omega)$ for single band is given by \cite{imada} :
\begin{equation}
A(\mathbf{k},\omega)=\frac{1}{\pi}\frac{\mid Im \Sigma(\mathbf{k},\omega)\mid}{[\omega-\varepsilon^{0}(\mathbf{k})-Re \Sigma(\mathbf{k},\omega)]^{2}+ [Im \Sigma(\mathbf{k},\omega)]^{2}},
\end{equation}
where, $\omega$ is real frequency, $\varepsilon^{0}(\mathbf{k})$ is energy of a single electronic state in the non-interacting system with crystal momentum $\mathbf{k}$. The chemical potential is taken to be zero in the above expression. The terms $Re \Sigma(\mathbf{k},\omega)$ and $Im\Sigma(\mathbf{k},\omega)$ are the real and imaginary part of the self-energy $\Sigma(\mathbf{k},\omega)$, respectively. The $Im \Sigma(\mathbf{k},\omega)$ gives broadening due to electron-electron interaction while, $Re \Sigma(\mathbf{k},\omega)$ gives energy shift due to interaction from $\varepsilon^{0}(\mathbf{k})$.  The temperature dependent electronic structure is studied by using Green's function $G$ defined in Matsubara-time domain which is called temperature Green's function method. Another approach is by using conventional statistical mechanical approach to calculate $G$. In the temperature Green's function method, the concept of temperature is introduced through Matsubara-time. This method is easy to adopt in self-consistent calculations \cite{matsu1,matsu2,matsu3,antiksnte}.

The imaginary part of electron and phonon self-energy due to the electron-phonon interaction is obtained in the EPW as below \cite{epw} :

The imaginary part of electron self-energy:
\begin{eqnarray}
\Sigma^{"}(\omega,T)=\pi \sum_{m\nu} \int_{BZ}\frac{d\mathbf{q}}{\Omega_{BZ}}\mid g_{mn,\nu}(\mathbf{k},\mathbf{q})\mid ^{2} \nonumber \\
\times \lbrace [ n_{\mathbf{q} \nu}(T)+f_{m\mathbf{k}+\mathbf{q} \nu}(T)]\delta(\omega - (\varepsilon_{m\mathbf{k}+\mathbf{q}}-\varepsilon_{F})+\omega_{\mathbf{q}\nu}) \nonumber \\
+ [ n_{\mathbf{q} \nu}(T)+ 1 -f_{m\mathbf{k}+\mathbf{q} \nu}(T)]\delta(\omega - (\varepsilon_{m\mathbf{k}+\mathbf{q}}-\varepsilon_{F})-\omega_{\mathbf{q}\nu}) \rbrace. \nonumber \\
\end{eqnarray}
The imaginary part of phonon self-energy:
\begin{eqnarray}
\Pi^{"}(\omega,T)=2\pi \sum_{m\nu} \int_{BZ}\frac{d\mathbf{q}}{\Omega_{BZ}}\mid g_{mn,\nu}(\mathbf{k},\mathbf{q})\mid ^{2} \nonumber \\
\times [f_{n\mathbf{k}}(T)-f_{m\mathbf{k}+\mathbf{q}}(T)]\delta(\varepsilon_{m\mathbf{k}+\mathbf{q}}-\varepsilon_{\mathbf{k}}-\omega).
\end{eqnarray}

In the above expressions $g_{mn,\nu}(\mathbf{k},\mathbf{q})$ is the first-order electron-phonon matrix elements from density functional perturbation theory (DFPT) calculations. This gives a quantification of the scattering process between Kohn-Sham states $m\mathbf{k}+\mathbf{q}$ and $n\mathbf{k}$ connected with a phonon of wave vector $\mathbf{q}$. $\omega_{\mathbf{q}\nu}$ is phonon frequency of branch index $\nu$. $\varepsilon_{n\mathbf{k}}$ is the eigenvalue of electron with band index $n$ and wave vector $\mathbf{k}$. $\varepsilon_{F}$ is the Fermi energy. The Bose-Einstein and Fermi-Dirac distributions are denoted by $n_{\mathbf{q}\nu}(T)$ and $f_{n\mathbf{k}}(T)$, respectively. A carrier lifetime (or relaxation time) due to EEI or EPI is obtained as $\hbar/linewidth$ \cite{landau}. Here linewidth is obtained as the twice of the imaginary part of electron or phonon self-energy.

The lifetime of phonons due to PPI is obtained from the imaginary part of phonon self-energy . The imaginary part of phonon self-energy of a mode, $\Gamma_{\lambda}(\omega)$ ($\lambda =\mathbf{q}\nu$) can be obtained using the third order anharmonic force constants under many-body perturbation theory as \cite{phono3py}: 

\begin{eqnarray}
\Gamma_\lambda(\omega)=\frac{18\pi}{\hbar^2}{\sum\limits_{\lambda^{\prime}\lambda^{\prime\prime} }}\mid \Phi_{-\lambda \lambda^\prime \lambda^{\prime\prime}}\mid^2 
 \{(n_{\lambda^\prime}+n_{\lambda^{\prime\prime}}+1) 
\nonumber \\
\times \delta(\omega-\omega_{\lambda^\prime}-\omega_{\lambda^{\prime\prime}})+(n_{\lambda^\prime}-n_{\lambda^{\prime\prime}}) 
\nonumber \\
\times [\delta(\omega+\omega_{\lambda^\prime}-\omega_{\lambda^{\prime\prime}}) \,- \, \delta(\omega-\omega_{\lambda^\prime}+\omega_{\lambda^{\prime\prime}})]  \}
\end{eqnarray}  
Here, $\Phi_{-\lambda \lambda^\prime \lambda^{\prime\prime}}$ represents strength of interaction between three phonon $\lambda$, $\lambda^{'}$ and $\lambda^{''}$ in the scattering. $\hbar$ is the reduced Planck's constant, $n_{\lambda}$ is the Bose-Einstein distribution function at equilibrium and $\omega_{\lambda}$ is frequency of phonon mode. 

Then, the lifetime (or relaxation time) of a mode $\lambda$ is obtained as $\tau_{\lambda}=1/2\Gamma_{\lambda}(\omega)$ \cite{phono3py}. Here, $2\Gamma_{\lambda}(\omega)$ is  called linewidth of a phonon mode $\lambda$.

\section{Computational details}
In this work, the electronic spectral function is calculated using full-$GW$ method in Matsubara time domain as implemented in Elk code \cite{elk}. A $\mathbf{q}$-mesh of size 4x4x4 and $\mathbf{k}$-mesh of size 8x8x8 are used in the calculation. The transformation of spectral function from imaginary to real axis is done under Pad\'{e} approximation \cite{pade}. For exchange-correlation (XC) LDA \cite{lda92} is used as starting point for the $GW$ calculation. The DFT band structure is obtained using WIEN2k \cite{wien2k} program. The electronic transport coefficients are calculated using BoltzTraP program \cite{boltztrap} combined with WIEN2k. The more details of these calculation can be found in Ref.\cite{shastri5}

In order to study EPI, the EPW program along with Quantum ESPRESSO (QE) package \cite{qe} is used. Using QE, initially the electronic scf and phonon (DFPT) calculations are carried out using project augmented wave (PAW) pseudopotential. For the XC part LDA is used. In electronic scf calculations a plane wave cutoff energy of 65 Ry is chosen. The convergence threshold for self-consistency is set to be 10$^{-10}$ Ry. A $\mathbf{k}$-point mesh of size 10 x 10 x 10 and 6 x 6 x 6 are used for scf and non-scf (nscf) calculations, respectively. The phonon calculations are carried out on 6x6x6 $\mathbf{q}$-mesh with convergence threshold of 10$^{-14}$ Ry. Next, the imaginary part of electron and phonon self-energy due to EPI is computed using the EPW program \cite{epw}. A finer $\mathbf{k}-$ and $\mathbf{q}-$mesh of 50 x 50 x 50 are used in the interpolation of electron-phonon matrix elements using Wannier functions. For phonon linewidth due to EPI calculation 23 x 23 x 23 sized $\mathbf{q}-$mesh is used. The Gaussian broadening parameter of 10 meV is used. The width of the Fermi surface window is taken to be 2 eV in the self-energy calculations. 

To calculate the lifetime of phonons due to PPI, phono3py \cite{phono3py}  combined with ABINIT package \cite{abinit} are used. Initially, the forces on atoms are obtained using the PAW method of DFT as implemented in the ABINIT package. To calculate the forces a supercell of the conventional unit cell of size 2 x 2 x 2 is built. The LDA is used for XC part. A plane wave energy cutoff of 25 Ha is used and twice of this value is used for the PAW energy cutoff. A force convergence criteria of 5x10$^{-8}$ Ha/bohr is used in scf calculations. A 4 x 4 x 4 k-mesh is used to sample the Brillouin zone of supercell. Next, the harmonic and anharmonic force constants are obtained under supercell method with a finite displacement of 0.06 bohr in phono3py \cite{phono3py}. The phonon properties and imaginary part of phonon self-energy due to PPI are obtained using a $\mathbf{q}$-mesh of size 23 x 23 x 23. The other details of $\kappa_{ph}$ calculations can be found in Ref. \cite{shastri5}.

\section{Results and Discussion}
 
\subsection{\label{sec:level2}Electron-electron interaction (EEI)}
The study of EEI and temperature effects on the electronic structure is important to understand the transport properties. Therfore, to see the effect of EEI and temperature on the electronic energy states of FeVSb, the spectral function $A_{j}(\mathbf{k},\omega)$ (here, $j$ is band index) is calculated at different  electronic temperatures using the full-$GW$ method. The obtained excited state electronic structure, $A_{j}(\mathbf{k},\omega)$ at the valence band maximum (VBM) at $L$-point ($L_{V}$) and conduction band bottom (CBM) at $X$-point ($C_{X}$) for four temperatures are shown in Fig. 1 (a). Since, the chemical potential in this $GW$ spectral function calculation is not determined, it is taken to be the middle of center of VBM peak P2 and CBM peak at 300 K for the reference purpose. In Fig. 1(a), the energy of $A_{j}(\mathbf{k},\omega)$ is shown with respect to the chemical potential (dotted line).  Here, the effect of EEI and temperature on $A_{j}(\mathbf{k},\omega)$ at the VBM and CBM are focused since, the major contribution to transport comes from the carriers in the neighborhood of these points. To see the changes in the electronic structure with the inclusion of EEI and temperature, the ground state electronic structure calculated using  DFT is also shown in Fig. 1 (b) for a reference purpose. It is known that in DFT, the interacting many particles are treated as non-interacting particles, but in an effective potential. As one can see from Fig. 1 (b), the VBM in the calculated DFT dispersion in FeVSb is doubly degenerate. But, as can be seen from Fig. 1 (a), this degeneracy is lifted and two well separated peaks (marked $P1$ and $P2$, in the order of energy) corresponding to VBM can be seen at 300 K. The energy difference between these two peaks is $\sim$91 meV. With the increase in temperature (300 K to 1000 K), the height of the peaks are observed to be reducing and simultaneously the broadening (width of peaks) is also increasing due to electronic excitations. At 1000 K, the separation between center of two peaks are very much reduced forming a hump-like shape. On observing the peaks' behaviour, it suggests that at still lower temperatures the separation between them becomes more prominent indicating a non-degeneracy at the VBM in FeVSb. Therefore, this result shows and suggests the importance of EEI in FeVSb.  Similarly, the height of the peak at CBM is found to be reducing and simultaneously broadening is increasing with the rise in temperature. Also, one can observe the effect of reduction in the energy separation between the VBM and CBM with the increase in temperature. Thus, to see this effect quantitatively we move on to estimate the temperature dependent band gap.  

\begin{figure*}
\includegraphics[width=14cm, height=6.5cm]{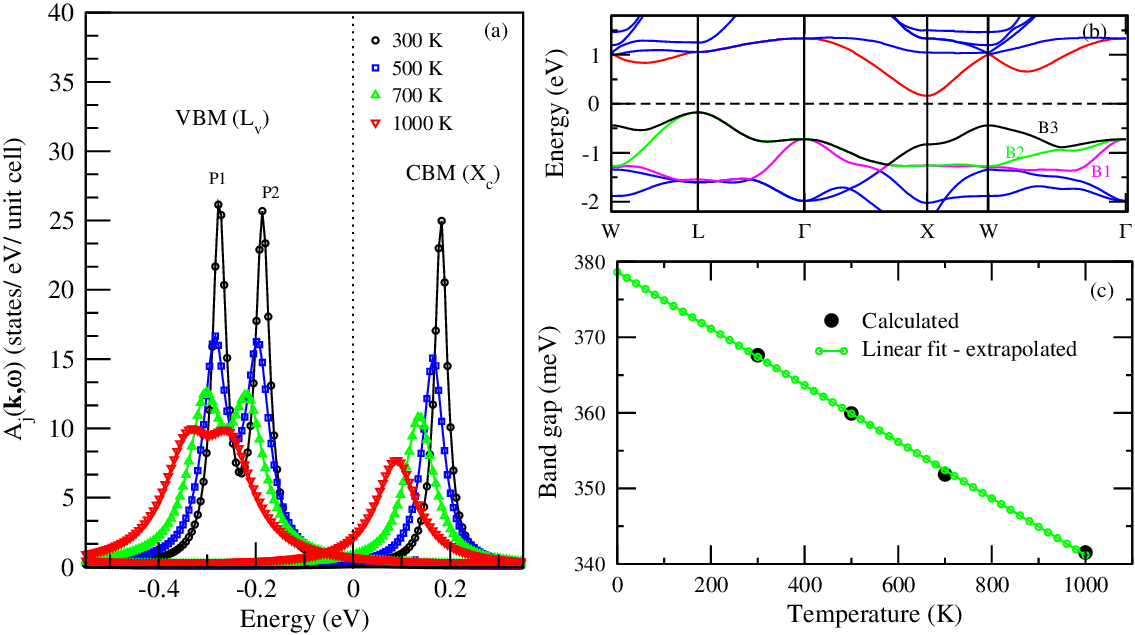} 
\caption{ (Colour online) (a) The spectral function at the $L$-point (VBM ($L_{v}$)) and $X$-point (CBM ($X_{c}$)) at different temperatures. (b) The ground state dispersion around the Fermi level. (c) The estimated variation of band gap from the temperature dependent spectral function.}
\end{figure*}

To see the effect of EEI and temperature on the electronic band gap (or more precisely quasi-particle gap), the energy separation between VBM and CBM is estimated from the $GW$ spectral function. Here, the indirect electronic gap is defined as the energy difference between the center of the peak $P2$ of VBM and the center of the peak of CBM at a given temperature. The calculated band gap with temperature is presented in Fig. 1 (c). The value of band gap at 300 K is $\sim$ 367 meV and it reaches a lower value of $\sim$ 341 meV at 1000 K. The calculated band gap data is fitted with linear equation and it is found to obey a linear decreasing trend with rise in temperature. In order to estimate ground state band gap considering the effect of EEI, the linear fit is extrapolated to 0 K. The obtained value of ground state band gap at 0 K is $\sim$378 meV. This ground state value of band gap is slightly higher than the band gap obtained from SCAN ($\sim$330 meV)  \cite{shastri5} and GGA-PBE ($\sim$320 meV) \cite{yangpbegap} DFT calculations. We could not come across any band gap from optical measurements in the literature for this compound. However, the thermal band gap estimated from Goldsmid and Sharp formula is 0.27 eV \cite{fubandgap}. This band gap is lower than the theoretically calculated band gaps which may be due to the defects or impurities present in the sample. The 300 K band gap is $\sim$3 \% reduced with respect to ground state band gap. Thus, the calculated spectral function shows the effect of temperature along with EEI on the electronic structure and band gap of FeVSb.

\begin{figure}
\includegraphics[width=5cm, height=4cm]{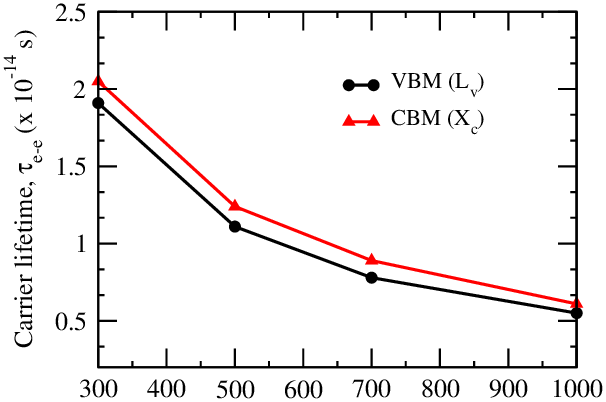} 
\caption{The lifetime of electrons at the VBM and CBM due to EEI with temperature.}%
\end{figure}  

The width of the quasiparticle (QP) peak gives information about the imaginary part of electron self-energy $\mid Im \Sigma(\mathbf{k},\omega)\mid$. This $\mid Im \Sigma(\mathbf{k},\omega)\mid$ is related to the QP-QP interaction. Using the $\mid Im \Sigma(\mathbf{k},\omega)\mid$ one can estimate the lifetime of carriers as mentioned in section-II. So, here the effect of EEI on the lifetime of particles at the band extrema are estimated from the $GW$-spectral function peaks. We calculated  $\mid Im \Sigma(\mathbf{k},\omega)\mid$ as the half-width at half maximum (HWHM)  of the Lorentzian fit to the peaks at VBM and CBM. Then, the lifetime of carriers ($\tau_{e-e}$) due EEI is calculated as $\hbar$/($2\times HWHM$). The obtained $\tau_{e-e}$  at the VBM and CBM of FeVSb for 300 K to 1000 K is shown in Fig. 2. The peaks $P1$ and $P2$ have the same width therefore same lifetimes are estimated for the particles. The $\tau_{e-e}$ has decreasing trend with the increase in temperature. Also, the particle lifetime at VBM is found to be slightly lower than that of CBM. At 300 K, the $\tau_{e-e}$ of particles at VBM (CBM) is $\sim$1.91 x10$^{-14}s$ ($\sim$2.05 x10$^{-14}s$). With the rise in temperature, the scattering increases and at 1000 K, the $\tau_{e-e}$ reaches a value of $\sim$0.55 x10$^{-14}s$ and $\sim$0.61 x10$^{-14}s$ at VBM and CBM, respectively. These calculated values of lifetime due to EEI suggests a high scattering of electrons in FeVSb at the band extrema. Also, the order of magnitude of lifetime suggests EEI is a major source of scattering in FeVSb and cannot be neglected in transport properties calculation. These estimated lifetime are due to the EEI but, the thermal vibration of lattice acts as another intrinsic source of scattering of electrons in crystalline solids. This interaction between electron and phonons plays important role in the transport mechanism which is studied for FeVSb in the next section.

\subsection{\label{sec:level2}Electron-phonon interaction (EPI)}
The thermal vibrations causes the ions/nuclei to undergo oscillatory motion from the equilibrium positions. This motion of nuclei affects the motion of electrons and vice-versa coupling to each other  which is called electron-phonon interaction (EPI). This effect of EPI on the lifetime of electrons and phonons is studied by calculating the electron and phonon linewidth in FeVSb. In Fig. 3 (a), the linewidth of electrons due to EPI at room temperature (300 K) along the high symmetric $\mathbf{k}$-directions in the Brillouin zone (BZ) are shown. The electron linewidths are calculated for the three bands (denoted as B1, B2 and B3 in Fig. 1 (b)) close to the valence band edge. The value of electron linewidth at the doubly degenerate VBM at $L$-point is $\sim$3.4 meV and lower compared to that of other high symmetric $\mathbf{k}$-points. At the $X$-point electron linewidth is $\sim$244 meV for band B3 which is higher compared to other $\mathbf{k}$-points shown. This indicates that the electrons in the neighborhood of VBM at $L$-point is less affected by the EPI and have lesser scattering (or more lifetime) relative to other $\mathbf{k}$-points at the valence band edge. 

\begin{figure*}
\includegraphics[width=14cm, height=5cm]{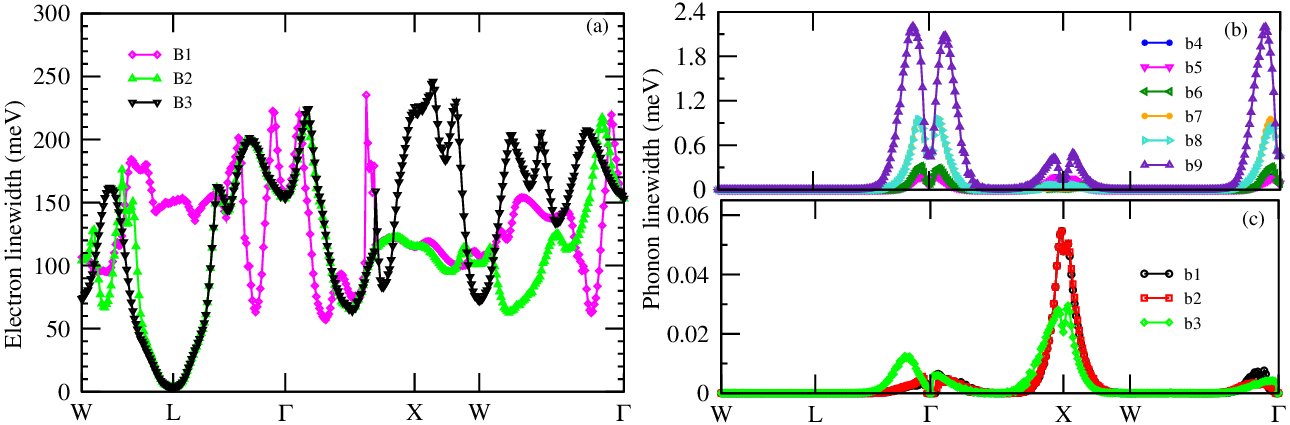} 
\caption{(a) The electron linewidth due to EPI at room temperature for three bands close valence band edge (marked B1, B2 and B3 in Fig. 1 (b)). (b) The optical phonons linewidth and (c) acoustic phonons linewidth due to EPI at room temperature.}
\end{figure*}

In order to study the effect EPI on the phonons, linewidth and lifetime are calculated for FeVSb half-Heusler. In our earlier work, we have calculated the phonon properties of FeVSb \cite{shastri5}. The phonon dispersion in FeVSb has six optical and three acoustic branches. For these phonon branches the linewidth and lifetime are calculated. The calculated room temperature linewidth of optical phonon branches (b4-b9) and acoustic phonon branches (b1-b3) along high symmetric $\mathbf{q}$-directions are shown in Fig. 3 (b) and (c), respectively to see the main scattering regions in the BZ. From the energy scale of acoustic and optical phonon linewidths, one can see that acoustic phonons have overall lower interactions with the valence band electrons compared to the optical phonons. This also suggests that the lifetime due to interactions with electrons for acoustic phonons is higher compared to optical phonons. The optical phonons are mainly scattered after interacting with electrons in FeVSb. From Fig 3.(b) one can see that the optical phonons close to $\Gamma$-point have higher linewidth relative to that of other high symmetric points. Also, one can observe that the highest energy branch (b9) optical phonons experience more scattering around the $\Gamma$-point with linewidth of $\sim$2.1 meV. The acoustic phonons have higher scattering around the $X$-point relative to other points as can be observed from Fig. 3(c) which is suggested by a linewidth of $\sim$0.05 meV for b1, b2 and $\sim$0.03 meV for b3. 

Further, to see the effect of temperature on the EPI and carrier scattering, the lifetimes of electrons and phonons are calculated at higher temperatures. The calculated electron lifetime $\tau^{e}_{e-ph}$ due to EPI, at the VBM at $L$-point in the temperature range 300 to 1200 K is shown in Fig. 4 (a) for FeVSb. At a given temperature, the value of $\tau^{e}_{e-ph}$ is obtained from the calculated imaginary part of electron self-energy due to EPI, using the formula mentioned before. The $\tau^{e}_{e-ph}$, has decreasing trend in the temperature range shown. The room temperature $\tau^{e}_{e-ph}$ is $\sim$19.3 x10$^{-14}s$ and it decrease to a value of $\sim$3.7 x10$^{-14}s$ at 1200 K. The values of $\tau^{e}_{e-e}$ is slightly lower compared to $\tau^{e}_{e-ph}$ indicating that effect of EEI is comparable to EPI at the VBM. This suggests that EEI is not ignorable in computing transport coefficients. We could not come across any reported value of electron lifetime due to EPI in literature for FeVSb. However, G. Samsonidze \textit{et al.} have studied p-type HfCoSb and n-type HfNiSn Heuslers with EPI \citep{kozinsky}. The reported value of electron lifetime at 673 K, for p-type HfCoSb at VBM is $\sim$20 x10$^{-15}s$ and for n-type HfNiSn at CBM is $\sim$80 x10$^{-15}s$. 

\begin{figure*}
\includegraphics[width=14cm, height=4.5cm]{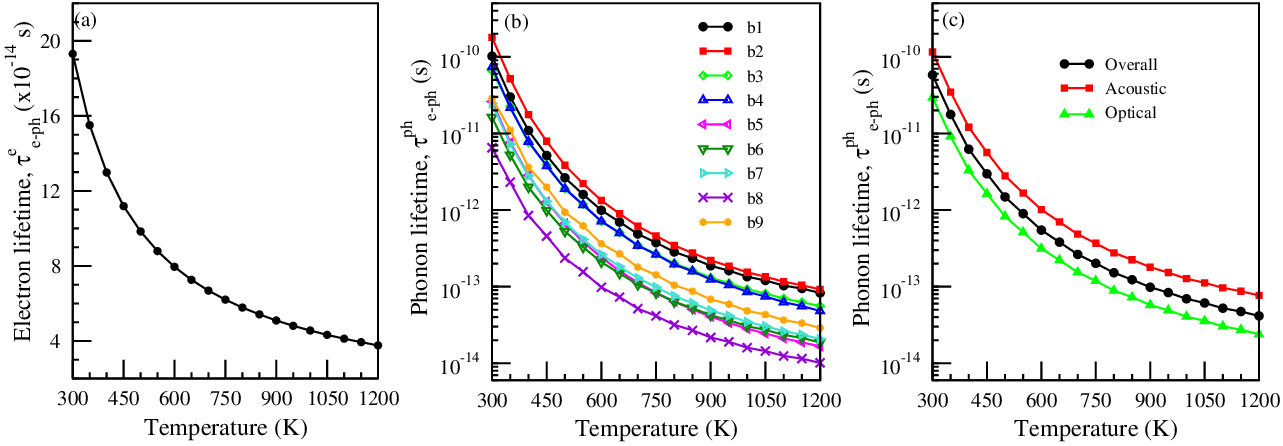} 
\caption{(a) The electron lifetime due to EPI, $\tau^{e}_{e-ph}$ at the VBM  as a function of temperature. (b) The phonon lifetime due to EPI for nine phonon branches.  (c) Acoustic, optical and overall phonon lifetime.}
\end{figure*}

The phonons in the solid also interact with electrons and have finite lifetime due to this interaction which is important in deciding the heat transport through them. Thus, to see the EPI effect on phonons, linewidth and lifetime at different temperatures are calculated for all points in irreducible part of BZ (IBZ). The phonon mode lifetime due to EPI is obtained as inverse of phonon mode linewidth  using the formula mentioned earlier. At a given temperature, the $\tau^{ph}_{e-ph}$, for a particular phonon branch is obtained as the weighted average over the $\mathbf{q}$-points in the BZ which is shown in Fig 4 (b) for 300 K - 1200 K. The branches b1-b3 are acoustic phonon branches and b4-b9 are optical branches. Further, acoustic, optical and overall phonon lifetimes are obtained by averaging over respective number of phonon branches. This is presented in Fig. 4 (c) as a function of temperature for FeVSb. From Fig. 4 (b), the range in values of lifetime for different phonon branches at a particular temperature can be observed. The figure also indicates order of magnitude changes in the $\tau^{ph}_{e-ph}$ with temperature. The lifetime of acoustic phonons shown in Fig. 4 (c), indicates that the optical phonons strongly interact with the electrons and thus have lower lifetime compared to acoustic phonons. This analysis suggests acoustic phonons mainly transport heat in FeVSb while optical phonons experience more scattering. The value of $\tau^{ph}_{e-ph}$ in FeVSb is reducing drastically (by 3 order of magnitude) from room temperature to 1200 K. The overall $\tau^{ph}_{e-ph}$ at 300 K is $\sim$5.8 x10$^{-11}s$ and reaches a value of $\sim$4.2 x10$^{-14}s$ at 1200 K. This large reduction in lifetime indicates that, EPI plays the major role in deciding the lattice thermal conductivity behavior. 

\subsection{\label{sec:level2}Phonon-phonon interaction (PPI)}
The phonons interact with other phonons in the solid. This phonon-phonon interaction (PPI) acts as another intrinsic source of phonon scattering leading to finite lifetime of phonons. In this section, the average phonon lifetime due to the PPI with temperature is calculated to understand $\kappa_{ph}$ behavior. The lifetime for each of phonon modes is calculated from phonon mode linewidth using the relation mentioned in section-II. Then for each of the nine phonon branches, the branch phonon lifetime $\tau^{ph}_{ph-ph}$ is calculated as the weighted average over the $\mathbf{q}$-points in the BZ. Further, acoustic, optical and overall phonon lifetimes $\tau^{ph}_{ph-ph}$ are obtained by averaging the branch lifetime over respective number of phonon branches at each temperature. These results calculated for the temperature range 300 K to 1200 K are shown in Fig. 5 (a) and (b), respectively.  
\begin{figure*}
\includegraphics[width=12cm, height=4cm]{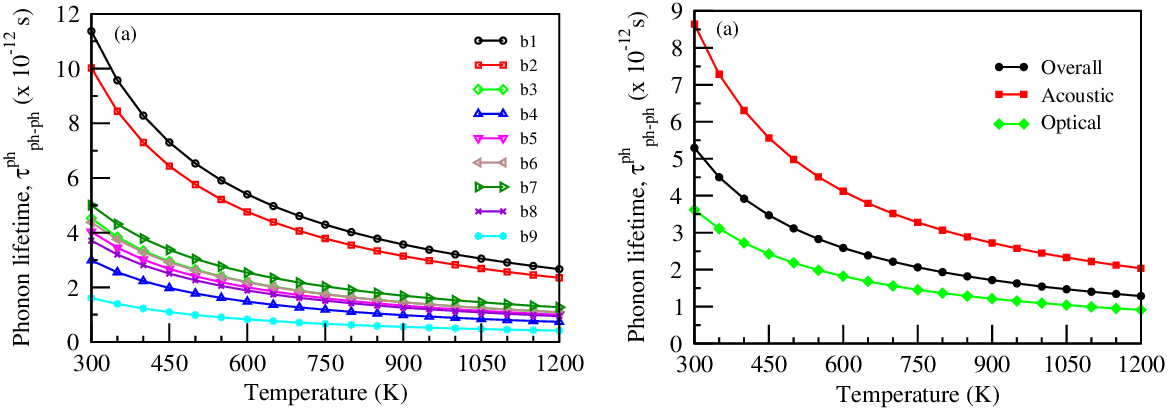} 
\caption{(a) The phonon lifetime for nine phonon branches. (b) Acoustic, optical and overall phonon lifetime due to PPI as a function of temperature in FeVSb.}
\end{figure*}

From Fig. 5 (a), one can observe that highest energy optical branch (b9) has the lowest lifetime indicating large scattering due to PPI. The acoustic branch phonons (b1 and b2) have higher lifetime and experience relatively lesser scattering. In Fig. 5 (b), the lifetimes of acoustic and optical phonons along with overall phonon lifetime as a function of temperature in FeVSb is presented. As can be seen from the figure, the nature of the phonon lifetime curve with temperature is similar to that of $\kappa_{ph}$ calculated in our previous work for FeVSb \cite{shastri5}. The group velocity is a temperature independent quantity in the method of calculation of $\kappa_{ph}$ used in Ref. \cite{shastri5} and specific heat is almost constant above $\sim$400 K. Therefore, the nature of phonon lifetime curve with temperature suggests that the lattice part of thermal conductivity due to PPI is mainly decided by the phonon lifetime. The value of overall $\tau^{ph}_{ph-ph}$ at room temperature is $\sim$5.3 x10$^{-12}s$ and it reaches to a value of $\sim$1.2 x10$^{-12}s$ at 1200 K. Also, it can be observed that the optical phonons have shorter lifetime compared acoustic phonons. Similar trend for lifetime of optical and acoustic phonons due to EPI can also be observed in Fig. 4 (c). This suggests that acoustic phonons undergo lesser scattering and carry more heat energy compared to optical phonons in FeVSb. The cumulative lattice thermal conductivity analysis carried out in our previous work also suggested that larger contribution to $\kappa_{ph}$ comes from acoustic phonons. The lifetime analysis in this work further supports this finding in FeVSb thermoelectric half-Heusler. Here, it is interesting to notice by comparing Fig. 4 (c) and 5 (b) that the values of overall $\tau^{ph}_{e-ph}$ are lesser compared to overall $\tau^{ph}_{ph-ph}$ above 500 K. However, the values of $\tau^{ph}_{ph-ph}$  are in the order of 10$^{-12}s$ in the temperature range studied. This suggests that at higher temperatures (above $\sim$550 K) EPI may be the dominating source of scattering compared to PPI for phonon in FeVSb. Also, the EPI is a major source of scattering in deciding both the electron and phonon transport and hence cannot be neglected in calculating or predicting transport properties.  
\begin{figure*}
\includegraphics[width=11cm, height=5cm]{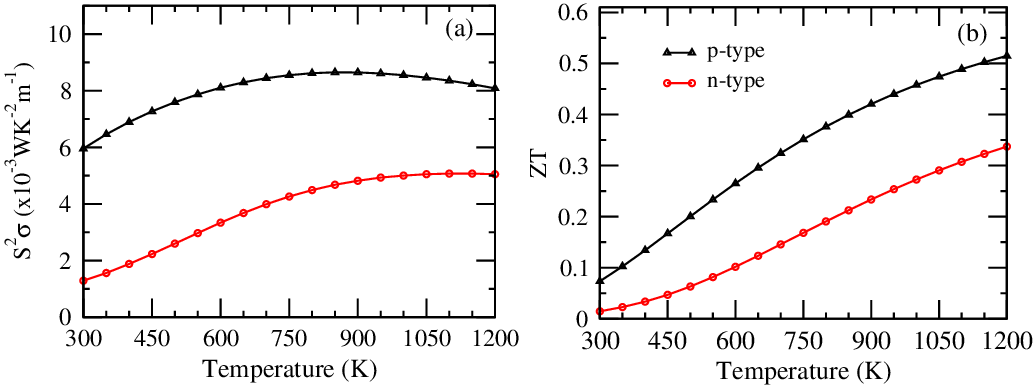} 
\caption{(a) Variation of power factor $S^{2}\sigma$ and (b) figure of merit $ZT$ with temperature T for p-type (triangle) and n-type (circle) FeVSb.}
\end{figure*}

\subsection{\label{sec:level2}Thermoelectric properties}
In this section, a prediction of power factor (PF) $S^{2}\sigma$ and figure of merit $ZT$ for p-type and n-type FeVSb using the first-principles carrier lifetimes is presented. In the BoltzTraP program which is normally used to calculate thermoelectric properties, the transport coefficients are obtained under constant relaxation time approximation (CRTA). But, the variation of carrier lifetime with temperature is one of the important quantities in deciding the transport behaviour which is not captured in such calculations. Therefore, here a prediction of thermoelectric properties is made by using the temperature dependent carrier lifetime estimated from first-principles, in the electronic transport coefficients calculated from BoltzTraP calculations in our last work \cite{shastri5}.

The calculated PF and $ZT$ for the p-type and n-type FeVSb as a function of temperature are shown in Fig. 6 (a) and (b), respectively. Here, the electronic transport coefficients appearing in $ZT$ are calculated by using temperature dependent carrier lifetimes ($\tau^{e}(T)$) obtained in this work. The carrier lifetime $\tau^{e}_{e-e}$ due to EEI at VBM and CBM are fitted with a function of the form $y=a\times x^{b}$ and extrapolated upto 1200 K. Using this fit, the values of $\tau^{e}_{e-e}$ at the intermediate temperature are obtained. The value of coefficient $a$ and $b$ for VBM (CBM) is 701.81 (631.96) and -1.0367 (-1.0038), respectively. The overall carrier lifetime at each temperature $\tau^{e}(T)$ is obtained from the $\tau^{e}_{e-e}(T)$ and $\tau^{e}_{e-ph}(T)$ using Matthiessen's rule. From Fig. 6, one can see that the PF and $ZT$ are higher for the p-type FeVSb compared to that of n-type FeVSb. The value of PF for p-type (n-type) FeVSb is $\sim$6 x10$^{-3} WK^{-2}m^{-1}$ ($\sim$1.3 x10$^{-3} WK^{-2}m^{-1}$) at 300 K and reaches a value of $\sim$8 x10$^{-3} WK^{-2}m^{-1}$ ($\sim$5 x10$^{-3} WK^{-2}m^{-1}$) at 1200 K. The PF for p-type FeVSb is increasing upto $\sim$800 K and above this temperature there is slow reduction in the value. Similarly, the PF for n-type FeVSb is increasing upto $\sim$900 K and then it is almost constant. The $ZT$ for the p-type (n-type) FeVSb at 300 K is 0.07 (0.01) and it reaches the maximum $ZT$ of 0.51 (0.34) at 1200 K. The predicted $ZT$ values suggest that the FeVSb can be used for thermoelectric application in the high temperature region. Also, this predicted values of $ZT$ give a more realistic result for FeVSb half-Heusler. In the calculation of $ZT$, $\kappa_{ph}$ values are used from our previous work \cite{shastri5}. Here, mode phonon lifetime due to EPI is not considered in the calculation of $\kappa_{ph}$, which can lead to further improvement in $ZT$. In this direction, a more detailed and careful study is needed to take into account mode wise phonon lifetime and band index, $\mathbf{k}$-point dependent electron lifetime in transport coefficient calculations.

\section{Conclusions} 
In summary, the lifetime of charge and heat carriers due intrinsic scattering mechanisms, $viz.$ electron-electron, electron-phonon and phonon-phonon interactions are studied from first-principles in FeVSb. The temperature dependent electronic structure at the VBM and CBM are studied by calculating the spectral function from the full-$GW$ method considering EEI. The results show a lift of degeneracy due to EEI at the VBM compared to DFT dispersion. The electronic band gap in FeVSb is found to decrease linearly with increase in temperature from this calculations. The obtained room temperature band gap with EEI is $\sim$ 367 meV and it reaches a lower value of $\sim$ 341 meV at 1000 K. The lifetime of carriers at VBM and CBM are estimated considering EEI with temperature. The room temperature value of lifetime at VBM (CBM) is $\sim$1.91 x10$^{-14}s$ ($\sim$2.05 x10$^{-14}s$). The values calculated suggest in fact, the EEI is of considerable importance in electronic transport. The effect of EPI is discussed by calculating temperature dependent lifetime of electrons and phonons. A branch wise analysis of lifetime of phonons due to EPI is carried out. The calculated values of electron lifetime suggest, both EEI and EPI should be taken into account in estimating the electronic transport parameters. Further, the effect of PPI on phonon lifetime is analysed by computing the lifetime for acoustic and optical phonon branches. This analysis suggest, the acoustic phonons experience lesser scattering due to PPI and EPI compared to optical phonons playing a role of major heat carriers. Also, the comparison of phonon lifetime due to EPI ($\sim$8.9 x10$^{-13}s$ at 550 K) and PPI ($\sim$2.8 x10$^{-12}s$ at 550 K) suggested, the EPI is the dominant scattering mechanism above 500 K for phonons and hence cannot be ignored in $\kappa_{ph}$ calculation and $ZT$ predictions. Lastly, a prediction of $ZT$ of p-type and n-type FeVSb is made by considering the temperature dependent carrier lifetime calculated for the electronic transport terms.

\section{Acknowledgements}
The authors thank Science and Engineering Research Board (SERB), Department of Science and Technology, Government of India for funding this work. This work is funded under the SERB project sanction order No. EMR/2016/001511.

\section{References}
\bibliography{ref_p6}
\bibliographystyle{apsrev4-1}

\end{document}